\documentclass[11pt,twoside]{article}  % Leave intact
\usepackage{apn3conf}

\begin{document}   % Leave intact

%-----------------------------------------------------------------------
%		            Paper Title 
%-----------------------------------------------------------------------
% Enter the title of the paper.
%
% EXAMPLE: \title{A Breakthrough in Astronomical Software Development}
% 
% If your title is so long as to fill the page header when you print it,
% then please supply a short form as a \titlemark.
%
% EXAMPLE: 
%  \title{Rapid Development for Distributed Computing, with Implications
%         for the Virtual Observatory}
%  \titlemark{Rapid Development for Distributed Computing}
%

\title{NGC\,1535 : UV Observations and Models}
%\titlemark{ }

%-----------------------------------------------------------------------
%		          Authors of Paper
%-----------------------------------------------------------------------
% Enter the authors followed by their affiliations.  The \author and
% \affil commands may appear multiple times as necessary (see example
% below).  List each author by giving the first name or initials first
% followed by the last name.  Authors with the same affiliations
% should grouped together. 
%
% EXAMPLE: \author{Margaret Meixner\altaffilmark{1}, Letizia Stanghellini,
%			Howard Bond} 
%          \affil{Space Telescope Science Institute, 
%                 3700 San Martin Dr.,  Baltimore, MD 21218}
%          \author{Joel Kastner}
%          \affil{Rochester Institute of Technology}
%
%          \altaffiltext{1}{Astronomy Department, UIUC}
%
% In this example, the first three authors, "Meixner", "Stanghellini", and
% "Bond" are affiliated with "STScI".  "Meixner" has an alternate 
% affiliation with the "Astronomy Department at UIUC".  The fourth author,
% "Kastner", is affiliated with "Rochester Institute of Technology"

\author{Lars Koesterke}
\affil{Laboratory for Astronomy and Solar Physics, NASA Goddard Space Flight Center,
         Greenbelt MD, USA}
\author{Klaus Werner}
\affil{Institut f\"{u}r Astronomie \& Astrophysics, University of T\"{u}bingen, Germany}
\author{Jeff W. Kruk}
\affil{Department of Physics \& Astonomy, JHU, Baltimore MD, USA}
\author{Thierry Lanz}
\affil{Laboratory for Astronomy and Solar Physics, NASA Goddard Space Flight Center,
         Greenbelt MD, USA}

%-----------------------------------------------------------------------
%			 Contact Information
%-----------------------------------------------------------------------
% This information will not appear in the paper but will be used by
% the editors in case you need to be contacted concerning your
% submission.  Enter your name as the contact along with your email
% address.
% 
% EXAMPLE:  \contact{Dennis Crabtree}
%           \email{crabtree@cfht.hawaii.edu}
%

\contact{Lars Koesterke}
\email{lars@winds.gsfc.nasa.gov}

%-----------------------------------------------------------------------
%		      Author Index Specification
%-----------------------------------------------------------------------
% Specify how each author name should appear in the author index.  The 
% \paindex{ } should be used to indicate the primary author, and the
% \aindex for all other co-authors.  You MUST use the following
% syntax: 
%
% SYNTAX:  \aindex{LASTNAME, F. M.}
% 
% where F is the first initial and M is the second initial (if
% used).  This guarantees that authors that appear in multiple papers
% will appear only once in the author index.  
%
% EXAMPLE: \paindex{Crabtree, D.}
%          \aindex{Manset, N.}        
%          \aindex{Veillet, C.}        
%
% NOTE: this information is also used to build the author list that
% appears in the table of contents.  Authors will be listed in the order
% of the \paindex and \aindex commmands.
%

\paindex{Koesterke, L.}
\aindex{Werner, K.}     % Remove this line if there is only one author
\aindex{Kruk, J. W.}     % Remove this line if there is only one author
\aindex{Lanz, T.}     % Remove this line if there is only one author

%-----------------------------------------------------------------------
%		      Author list for page header	
%-----------------------------------------------------------------------
% Please supply a list of author last names for the page header. in
% one of these formats:
%
% EXAMPLES:
% \authormark{LASTNAME}
% \authormark{LASTNAME1 \& LASTNAME2}
% \authormark{LASTNAME1, LASTNAME2, ... \& LASTNAMEn}
% \authormark{LASTNAME et al.}
%
% Use the "et al." form in the case of seven or more authors, or if
% the preferred form is too long to fit in the header.

\authormark{Koesterke, Werner, Kruk \& Lanz}

%-----------------------------------------------------------------------
%			Subject Index keywords
%-----------------------------------------------------------------------
% Enter up to 6 keywords describing your paper.  These will NOT be
% printed as part of your paper; however, they will be used to
% generate an object index and a subject index for the proceedings.  
% There is no standard list,  however, individual object names are
% encouraged and one or two word descriptions of the topics (e.g.MHD, 
% ionized gas) are useful. 
%
% EXAMPLE:  \keywords{NGC 7027, AFGL 2688, HD 161796, binary stars,
%                      dust,  molecular gas}
%


%-----------------------------------------------------------------------
%			       Abstract
%-----------------------------------------------------------------------
% Type abstract in the space below.  Consult the User Guide and Latex
% Information file for a list of supported macros (e.g. for typesetting 
% special symbols). Do not leave a blank line between \begin{abstract} 
% and the start of your text.

\begin{abstract}          % Leave intact
% Place the text of your abstract here - NO BLANK LINES
We reinvestigate the UV spectrum of NGC\,1535 by means of recently
developed fully line-blanketed non-LTE models. These new models account
for the wind in spherical geometry while handling the atomic data in a
very similar way to the {\sc Tlusty} code. This approach ensures at the same
time realistic predictions of the photospheric absorption lines and of the
emission lines formed in the wind. 
Our analysis confirms the results of previous studies. We derive
$T_*=70\,{\rm kK}$,
{\raisebox{0.24ex}{$\stackrel{{\textstyle\raisebox{-0.2ex}{.}}}{M}$}}
$=10^{-7.8}\,{\rm M_\odot/yr}$, and $v_\infty$=2000\,km/s.
\end{abstract}

%-----------------------------------------------------------------------
%			      Main Body
%-----------------------------------------------------------------------
% Place the text for the main body of the paper here.  You should use
% the \section command to label the various sections; use of
% \subsection is optional.  Significant words in section titles should
% be capitalized.  Sections and subsections will be numbered
% automatically. 
%
% EXAMPLE:  \section{Introduction}
%           ...
%           \subsection{Our View of the World}
%           ...
%           \section{A New Approach}
%
% It is recommended that you look at the sample papers, sample1.tex
% and sample2.tex, for examples for formatting references, footnotes,
% figures, equations, html links, lists, and other special features.  

\section{Introduction}
NGC\,1535 is a planetary nebula with a hot central star which
belongs to the group of Hydrogen-rich objects with weak winds. The
stellar parameters and abundances of these stars can be determined 
from plane-parallel, static models by fitting their stellar absorption
lines (c.f. Werner et al. 2003). A small subset of this group, to which
NGC\,1535 belongs, shows P-Cygni profiles in the UV which allows for an
investigation of their weak stellar winds by means of spherical wind
models.

The stellar parameters and the wind properties of NGC\,1535 have been well
established by Mend\'{e}z et al. (MKH, 1992), Perinotto (P, 1993), Tinkler \&
Lamers (TL, 2002) and Bauer \& Husfeld (BH, 1995). In the present paper we
confirm these results by applying (for the first time) a new non-LTE
code for expanding stellar atmospheres to recently observed HST and FUSE
spectra.

\section{Observations}
We have taken new UV and far-UV spectra of NGC 1535 with HST and FUSE. The HST
observation was performed on 2001-03-01 using STIS with the E140M grating,
for a total exposure time of 2346\,s. It covers the wavelength range
1150--1730\AA\ at a resolution somewhat higher than 0.1\AA. The FUSE
observation with an exposure time of 7862\,s was performed on 2003-01-01
with a similar spectral resolution, covering the range 905--1185\AA.

\section{Model Calculations}
Over the recent years we have analyzed the spectra of several CSPN
by applying two non-LTE codes (c.f. Koesterke \& Werner 1998). The
T\"{u}bingen code (Rauch \& Werner 1997) was used to fit the
photospheric absorption lines while the Potsdam-Kiel code (Koesterke \&
Hamann 1997) was used to fit the winds lines. The reason behind that was
the fact that the Potsdam-Kiel code did not allow for the proper
calculation of the quasi-static photosphere. To overcome this unsatisfying situation
we decided to write a new non-LTE code which is applied here for the
first time. This new code accounts for the stellar wind while handling the
atomic data in a very similar way to the widely-used code {\sc Tlusty} (Hubeny \&
Lanz 1995), i.e. incorporating in particular iron line-blanketing
and detailed line broadening.

\section{Results}
We present three models for NGC\,1535 which differ in temperature and/or
mass-loss rate. The model atmospheres have a solar composition and include
H, He, C, N, O, Ne, Si, S, Fe and Ni. A summary of the derived
stellar parameters comapred to the results of previous studies
is given in Table\,1. 

The relatively strong P-Cygni profiles of the resonance lines of O\,{\sc
VI}, O\,{\sc V} and N\,{\sc V}, the weak wind line of
C\,{\sc IV} and the absorption line of He\,{\sc II} (Fig.\,1) are well
matched by the models. The observed flux is consistent with a reddening
of ${\rm E_{B-V}\,=\,0.05}$.

\begin{table}[hbtp]
%\scriptsize
%\centering
\caption[]{Parameters of NGC\,1535}
\medskip
\begin{tabular}{lcccccc}
Study$^a$& T$_*$& log g& 
M$_*$& log (L$_*$/L$_\odot$)&
log{\raisebox{0.24ex}{$\stackrel{{\textstyle\raisebox{-0.2ex}{.}}}{M}$}}&
v$_\infty$ \\
& kK& (cgs)& M$_\odot$& & M$_\odot$/yr& km/s \\
\hline
MKH& 70& 4.65& 0.65& 3.94& -& - \\
P& 77& -& 0.55& 3.98& -8.85& 1900 \\
TL& 80.8& 5.05& 0.605& 3.76& -8.7& 1900 \\
BH& 70& 4.6& 0.66& 4.0& -& - \\
This study& 70& 4.6& -& -& -7.8& 2000 \\
\hline
\end{tabular}
\\
$^a$see 2$^{\rm nd}$ paragraph of introduction
\end{table}

%---------------- Figure sp ---------------------------------------------
\begin{figure}[h!btp]
\centering
\epsfxsize=10.5cm
\mbox{\epsffile{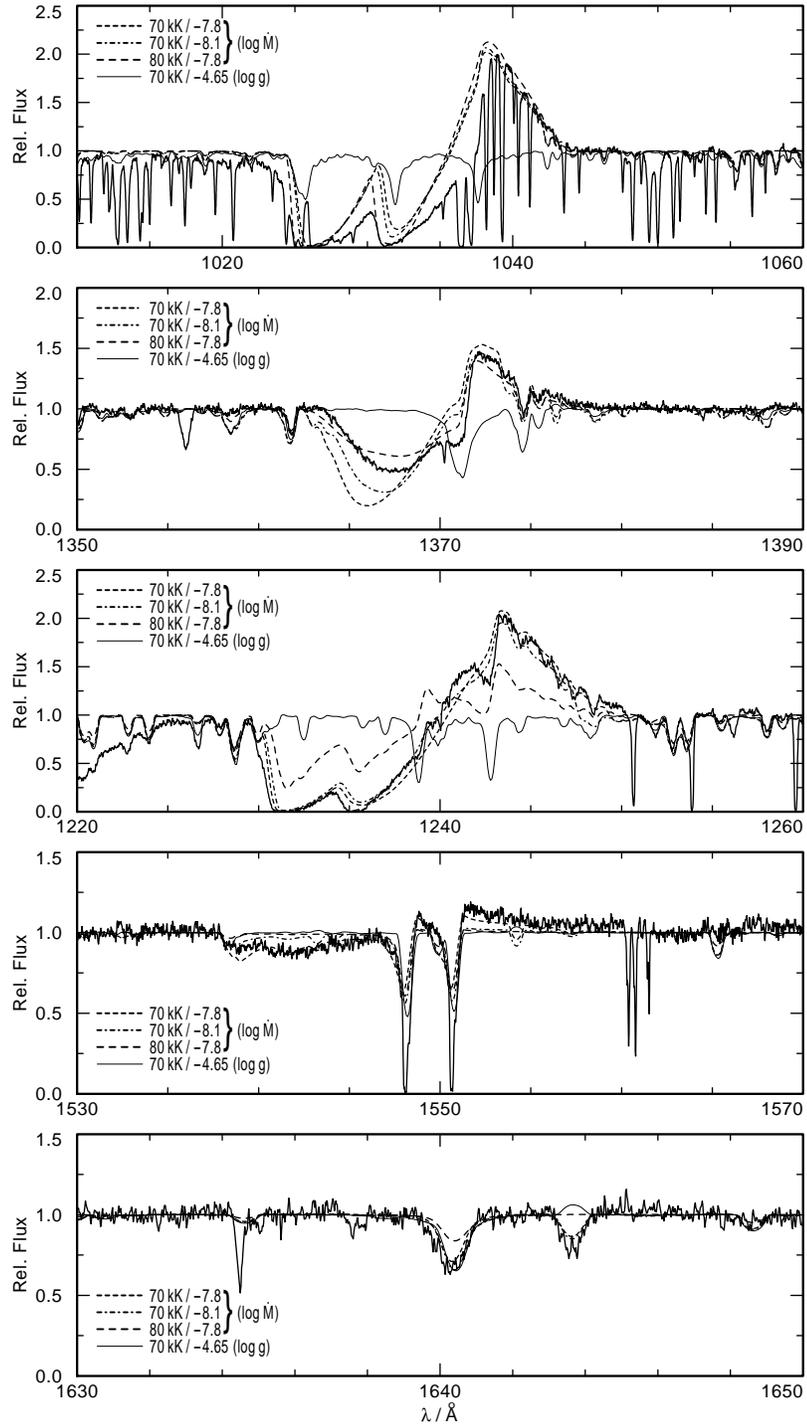}}
\caption [ ]{
Observation (solid line) of the UV lines compared
to the three Wind models (dashed, dash-dotted) and the {\sc Tlusty}
model (thin solid). A rotation velocity of 75\,km/s is adopted. 
}
\label{fig_dif1}
\end{figure}
%----------------------------------------------------------------------

%%---------------- Figure sp1 ---------------------------------------------
%\begin{figure}[h!btp]
%\centering
%\epsfxsize=9.5cm
%\mbox{\epsffile{sp1.ps}}
%\caption [ ]{
%Observation (solid line) of the He\,{\sc II} line at 1640\,\AA\ compared
%to the three Wind models (dashed, dash-dotted) and the {\sc Tlusty}
%model (thin solid). A rotation velocity of 75\,km/s is adopted. 
%}
%\label{fig_dif1}
%\end{figure}
%%----------------------------------------------------------------------

%%---------------- Figure sp2 ---------------------------------------------
%\begin{figure}[h!btp]
%\centering
%\epsfxsize=9.5cm
%\mbox{\epsffile{sp2.ps}}
%\caption [ ]{
%Observation (solid line) of the He\,{\sc II} line at 1640\,\AA\ compared
%to the three Wind models (dashed, dash-dotted) and the {\sc Tlusty}
%model (thin solid). A rotation velocity of 75\,km/s is adopted. 
%}
%\label{fig_dif1}
%\end{figure}
%%----------------------------------------------------------------------

%---------------- Figure 6 ---------------------------------------------
\begin{figure}[hbtp]
\centering
\epsfxsize=13.2cm
\mbox{\epsffile{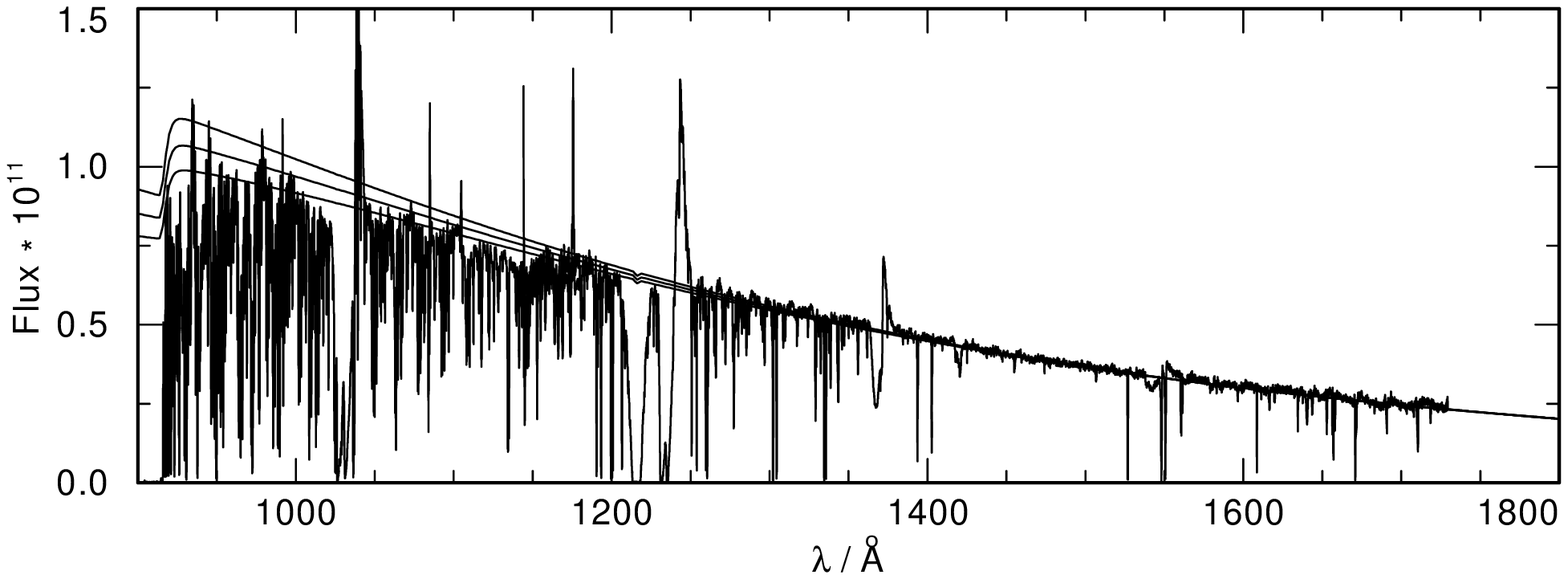}}
\caption [ ]{
Stellar flux compared to model flux (T\,=\,70\,kK, 
log{\raisebox{0.24ex}{$\stackrel{{\textstyle\raisebox{-0.2ex}{.}}}{M}$}}\,=\,
-7.8\,[M$_\odot$/yr]). The model flux is reddened by ${\rm E_{B-V}}$=0.04,
0.05 and 0.06, respectively and normalized to the observation at
1500\AA. 
}
\label{fig_dif1}
\end{figure}


\begin{references}
\reference Bauer F., Husfeld D. 1995, \aap\ 300, 481
\reference Hubeny I., Lanz T. 1995, \apj\ 439, 875
\reference Koesterke L., Hamann W.-R. 1997, \aap\ 320, 91
\reference Koesterke L., Werner K. 1998, \apj\ 500, L55
\reference M\'{e}ndez R.H., Kudritzki R.P., Herrero A. 1992, \aap\ 260, 329
\reference Perinotto M. 1993, in Planetary Nebulae, ed. R. Weinberger \&
           A. Acker, IAU Symp. 155, p57
\reference Rauch T., Werner K. 1997 in The Third Conference on Faint
           Blue Stars eds. A.G.D. Philip, J. Liebert, R. Saffer,
	   D.S. Hayes, L. Davis Press, p217
\reference Tinkler C.M., Lamers H.J.G.L.M. 2002, \aap\ 384, 987
\reference Werner K., Deetjen J.L., Dreizler S., Rauch T. 2003, in
           Planetary Nebulae: Their Evolution and Role in the Universe,
	   ed. M. Dopita, ASP Conference Series

\end{references}
\end{document}